\documentclass[twocolumn,nofootinbib,amsmath,amssymb,a4paper]{revtex4}

\usepackage{graphicx}
\usepackage{dcolumn}
\usepackage{bm}
\usepackage{subfigure}

\begin{document}

\title{Threshold Resummation in Semi-Inclusive $B$ decays}

\author{Giancarlo Ferrera}
\affiliation{
Universitat de Barcelona, 
Barcelona,  
Spain
\&
Universidad de Granada, 
Granada, Spain
}
\begin{abstract}
We discuss threshold resummation in radiative and charmless semileptonic 
$B$ decays.
To deal with the large
non perturbative effects, we introduce a model for NNLL resummed form factors
based on the analytic QCD coupling.
By means of this model 
we can reproduce with good accuracy the experimental data. 
Finally we briefly present an 
improved threshold resummed formula to deal with jets
initiated by massive quarks as in the case of semileptonic charmed decays.
\end{abstract}

\maketitle

\section{Introduction}
The aim of the work presented in this talk is to analyze 
semi-inclusive $B$ decays spectra 
measured at $B$-factories, which allows the extraction
of the 
Cabibbo--Kobayashi--Maskawa (CKM) matrix elements $|V_{ub}|$ and $|V_{cb}|$
~\cite{cabibbo}, \cite{km}.
To this end  is crucial to have a good control of the so called 
threshold region, defined
as the region where the invariant mass of the inclusive hadronic state
$X$ is much smaller compared with its energy: $m_X \ll E_X$.
This region is affected both by perturbative soft gluon radiation 
and by non-perturbative phenomena related to the ``Fermi-motion'' of the
heavy-quark inside the meson~\cite{uraltsev}.
We take in account such phenomena
with a model based on soft gluon resummation to 
next-to-next-to-leading logarithmic accuracy (NNLL) and on analytic 
QCD coupling having no Landau pole.
Our model, which does 
not contain non-perturbative free-parameters, gives a good
description of experimental data of the $B$-factories 
and it allows an extraction of $\alpha_S(m_Z)$ which is in agreement 
with the current PDG average within at most two standard deviations.

\section{Threshold Resummation}
Let's consider first radiative decays $B \to X_s\,\gamma$: factorization
and resummation of threshold logarithms in such decays leads to an expression 
for the event fraction of the form
\vspace*{-.1cm}
\begin{equation}
\frac{1}{\Gamma_r}\int_0^{t_s}\frac{d\Gamma_r}{dt}dt = C_r[\alpha_S(Q)]
\,\Sigma[t_s;\alpha_S(Q)]
+ D_r[t_s;\alpha_S(Q)]\,, 
\vspace*{-.0cm}
\end{equation}
where $\Gamma_r$ is the inclusive radiative width,
$t_s\equiv m_{X_s}^2/m_b^2$, $C_r(\alpha_S)$ is a short-distance, process
dependent hard factor, 
$\Sigma(t_s,\alpha_S)$ is the universal QCD form factor for
heavy flavor decays resumming series of  
logarithmically enhanced terms to any order in $\alpha_S$
and $D_r(t_s,\alpha_S)$ is a short-distance, process
dependent remainder function vanishing in the threshold region $t_s\to 0$.

An analogous formula can be written for the semileptonic decays 
$B \to X_u\,l\,\nu_l$. In this three-body decay case, the most general
distribution is a triple differential distribution~\cite{ugo2001} 
\vspace*{-.1cm}
\begin{eqnarray}
\label{tripla}
\frac{1}{\Gamma_s} \int_0^u \frac{d^3\Gamma_s}{dx dw du'} du' 
&=& C_s[x,w;\alpha_S(Q)] 
\,\Sigma[u;\alpha_S(Q)]\nonumber\\
&+& D_s[x,u,w;\alpha_S(Q)],
\vspace*{-.1cm}
\end{eqnarray}
where
$x =  \frac{2 E_l}{m_b},~
w  =  \frac{2E_X}{m_b},~
u = \frac{1 - \sqrt{1 - (2m_X/Q)^2} }{1 + \sqrt{1 - (2m_X/Q)^2} }~.
$

The hard scale of the $B$ decays in the threshold region 
is fixed by the final hadronic energy
$E_X$
\vspace*{-.15cm}
\begin{equation}
Q = 2E_X = m_b\left(1-\frac{q^2}{m_b^2}+\frac{m_X}{m_b^2}\right),
\vspace*{-.15cm}
\end{equation}
where $q^{\mu}$ is the 4-momentum of the probe 
(the real photon in the
radiative decays and the lepton neutrino pair in the semileptonic ones).
While in the radiative decays $q^2 = 0$ and the hard scale
is always of the order of the beauty mass
($Q \simeq m_b $),
in the semileptonic decays the lepton pair can have a large
invariant mass $q^2 \sim m_b^2$, 
implying a substantial reduction of the hard scale,
which is no more fixed but it depends on the kinematics. 
Since the hard scale $Q$ appears in the argument of the infrared logarithms
as well as in the argument of the running coupling, 
semileptonic spectra have in general 
a specific infrared structure, which is different 
from the invariant hadron mass distribution in the 
radiative decay. 
Semileptonic spectra are naturally classified depending on
whether they involve or do not involve an integration over the hard scale. 
The main consequence is that additional long distance effects,
related to small hadronic energies, occur in semileptonic decays,
which cannot be extracted from the radiative decays~\cite{noi}.

The heavy flavor form factor has an exponential form in Mellin 
 space~\cite{cattren}, \cite{sterman}:
\vspace*{-.2cm}
\begin{eqnarray}
\label{expoGN}
\log \sigma_N(\alpha_S)
\!\!&=&\!\!\!\! 
\int_0^1 \!\!\frac{dy}{y} [ (1-y)^{N-1}\! - 1 ]
\Bigg\{\!\!
\int_{Q^2 y^2}^{Q^2 y} \!\!\frac{dk_{\perp}^2}{k_{\perp}^2}
\!{A}[{\alpha_S}(k_{\perp}^2)]\nonumber\\
 \!&+&\!  {B}[{\alpha_S}(Q^2 y)]
+  {D}[{\alpha_S}(Q^2 y^2)]
\Bigg\},
\vspace*{-.2cm}
\end{eqnarray}
\vspace*{-.4cm}\\
where 
$\sigma_N(\alpha_S) \!=\!  \int_0^1  (1-t)^{N-1}  \sigma(t;\alpha_S) dt$,
$\sigma(t,\alpha_S) = {d\Sigma(t,\alpha_S)}/{dt}$
and the functions ${A}({\alpha_S}),\, {B}({\alpha_S})$ and 
${D}({\alpha_S})$ describe log-enhanced radiation and have a standard 
fixed order expansions in $\alpha_S$.

As is clear from the $k_\perp$ integral of 
Eq.(\ref{expoGN}), semi-inclusive processes are multi-scale
processes, characterized by fluctuations with transverse momenta up to $Q$:
a jet with a relative large invariant mass $m_X$ 
--- typically $\Lambda_{QCD}\ll m_X \ll Q$ ---
can contain very soft partons, with transverse
momenta of the order of the hadronic scale. 
That produces an ill-defined integration over the Landau pole
and the form factor acquires an unphysical
imaginary part for any $N$.
A prescription for the low-energy behaviour of the running coupling is
therefore needed to give a meaning to the formal expression in 
Eq.(\ref{expoGN}): our prescription is to use an effective QCD coupling
which does not present the Landau pole. 

\section{Effective coupling}

The standard QCD coupling
\vspace*{-.2cm}
\begin{equation}
\alpha_{S}^{lo}(Q^2) = \frac{1}{\beta_0 \log(Q^2/\Lambda_{QCD}^2)},
\vspace*{-.2cm}
\end{equation}
where $Q^2 \equiv - q^2$ with $q^{\mu}$ the gluon momentum,
has a physical cut for $Q^2<0$ related to the decay 
of a time-like gluon in secondary partons ($g^* \, \to gg, \, q\bar{q}$, etc.) 
and an unphysical simple pole (the Landau pole) for $Q^2 = \Lambda_{QCD}^2$, 
which signals a formal breakdown of the perturbative scheme.

The analytic QCD coupling is defined as having the same discontinuity of the
standard coupling along the cut, while being analytic elsewhere in the complex
plane~\cite{shirkov},
\vspace*{-.2cm}
\begin{equation}
\label{disp_rel}
\bar{\alpha}_S(Q^2)  = 
\frac{1}{2\pi i}
\int_0^{\infty}   \frac{ds}{s  +  Q^2} 
{\rm Disc}_s  \alpha_S(-s).
\vspace*{-.2cm}
\end{equation}
At LO, it reads
\vspace*{-.2cm}
\begin{equation}
\label{end}
\bar{\alpha}_{S}^{lo}(Q^2)  =  \frac{1}{\beta_0}
\left[
\frac{ 1 }{ \log Q^2/\Lambda^2 }
 -  \frac{ \Lambda^2 }{ Q^2 - \Lambda^2 }
\right].
\end{equation}
\begin{figure}[h]
\vspace*{-.3cm}
\includegraphics[width=0.38\textwidth]{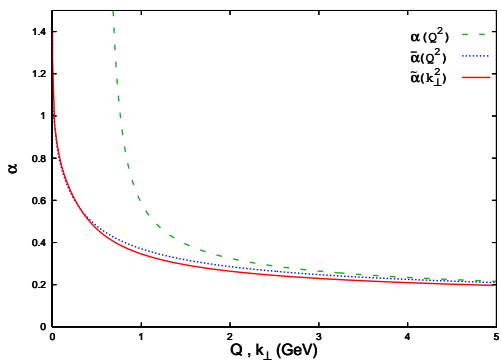}
\vspace*{-.1cm}
\caption{
QCD couplings at NNLO. 
Dashed line (green): standard coupling $\alpha_S(Q^2)$; dotted line (blue):
analytic coupling
$\bar{\alpha}_S(Q^2)$; continuous line (red):
time-like coupling $\tilde{\alpha}_S(k_{\perp}^2)$.
\label{coup}}
\vspace*{-.3cm}
\end{figure}

The analyticization procedure has therefore 
the effect of subtracting the infrared pole in $Q^2 \, = \, \Lambda^2$,
by means of a power-suppressed term,
in a minimal way. 
The analytic coupling (Fig.~\ref{coup})
has a constant limit at zero momentum transfer:
$\lim_{Q^2 \rightarrow 0} \bar{\alpha}_{S}(Q^2) =  \frac{1}{\beta_0}$);
furthermore, the subtraction term 
does not modify the high-energy behavior because it decays 
as an inverse power of the hard scale:
$ \bar{\alpha}_{S}(Q^2) \simeq \alpha_{S}(Q^2)$ for
$Q^2 \gg \Lambda_{QCD}$.

Higher orders in the form factors
have the main effect of replacing
the tree-level coupling with a time-like one evaluated at the transverse
momentum of the primary emitted gluon
\vspace*{-.35cm}
\begin{equation}
\alpha_S \,\,\, \to \,\,\,
\label{deftime}
\tilde{\alpha}_S(k_{\perp}^2) 
 \equiv  \frac{i}{2 \pi}  \int_0^{k_{\perp}^2}  d s
\, {\rm Disc}_s  \frac{ \alpha_S(-s) }{ s }.
\vspace*{-.1cm}
\end{equation}
By performing such integral exactly,
one includes in the coupling absorptive
effects related to the decay of time-like gluons.

The prescription at the root of our model is simply 
to replace the standard coupling on the r.h.s.\ of
Eq.~(\ref{deftime}) with the analytic 
coupling~\cite{ugogiu2004} (see also~\cite{stefanis}).
Therefore we have
a formula similar to 
Eq.(\ref{expoGN}) where the effective coupling have replaced the standard
one. 
In order to include as many corrections as possible,
we make the integration over $y$ in Eq.(\ref{expoGN}) 
exactly, in numerical way;
this is possible because the time-like coupling $\tilde{\alpha_S}(k_{\perp}^2)$
does not have the Landau singularity and is regular
for any $k_{\perp}^2 \, \ge \, 0$.

The form factor in the physical space is obtained by inverse Mellin transform
\vspace*{-.1cm}
\begin{equation}
\label{inverse}
\sigma(t; \, \alpha_S) = 
\int_{C-i\infty}^{C+i\infty} \frac{dN}{2\pi i}
(1-t)^{-N} \, \sigma_N(\alpha_S),
\vspace*{-.1cm}
\end{equation}
where the constant $C$ is chosen so that the integration contour
in the $N$-plane lies to the right of all the singularities of 
$\sigma_N(\alpha_S)$.
In order to correctly implement multi-parton kinematics, the inverse transform
from $N$-space back to $x$-space is also made exactly, in numerical way.
Let us note that no prescription --- such as the minimal prescription in the 
standard formalism \cite{mp} --- is needed in our model because 
$\sigma_N(\alpha_S)$ is analytic
for any ${\rm Re} \, N \, > \, 0$.

\section{phenomenology}

In this section we compare the theoretical distributions
obtained with our model with
experimental data (see also~\cite{modfrag}, for other approaches see
\cite{gardi}, \cite{neubert}, \cite{ligeti}).
Since our model has no free parameters, it allows a straightforward 
extraction of the value of $\alpha_S\left(m_Z\right)$ from the experimental 
data.
\begin{figure}[h]
\vspace*{-.16cm}
\includegraphics[width=0.38\textwidth]{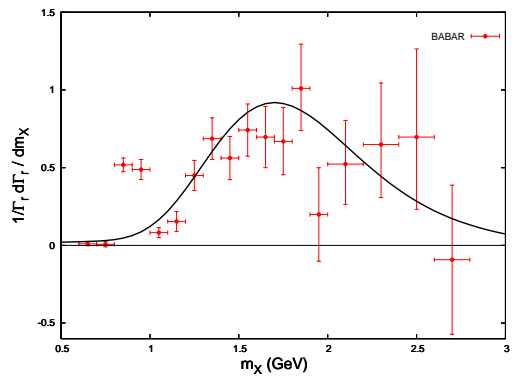}
\vspace*{-.1cm}
\caption{
$B\to X_s\,\gamma$\,: $m_X$
distribution from BaBar~\cite{babar}. 
The data show a rather pronounced $K^{*}$ peak, 
which clearly cannot be accounted for
in a perturbative QCD framework.
}
\label{rdmx}
\vspace*{-.75cm}
\end{figure}
\\
 The electron spectrum and the $m_X$ spectrum in the semileptonic
decay are affected by a large background for
$E_e  <   \frac{m_B}{2}  ( 1  - {m_D^2}/{m_B^2} )
 \simeq  2.31  {\rm GeV}$ (i.e.\ for $\bar{x}_e \, > \, 0.125$)
and for $m_X > m_D =1.867 GeV$
respectively, coming from the decays $B  \to  X_c\,l\,\nu_l$.
This background is larger than the signal by two orders of magnitude
because $|V_{ub}|^2/|V_{cb}|^2 \sim 10^{-2}$.
Let us stress that the photon energy spectrum in radiative decay
and the electron energy spectrum in semileptonic decay
have\, to\, be\, convoluted\, with\, a\, normal\, distribution,\; in\, order\, 
to\hspace*{-.15cm}
\begin{figure}[h]
\vspace*{-.1cm}
\includegraphics[width=0.38\textwidth]{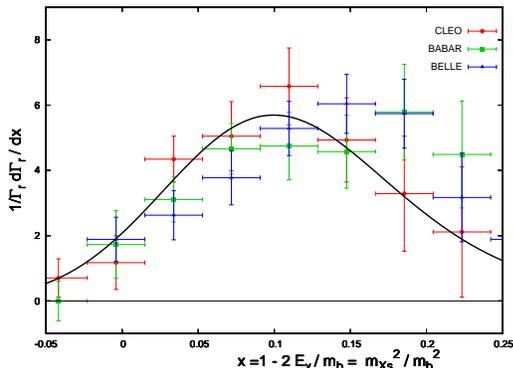}
\vspace*{-.2cm}
\caption{
\small
$B\to X_s\, \gamma$: photon spectrum from CLEO (red), BaBar (green) and Belle
(blue)~\cite{cleo,babar, belle}, 
the Doppler effect is sufficient to completely eliminate
the $K^*$ peak.
}
\label{ef}
\vspace*{-.1cm}
\end{figure}
\begin{figure}[h]
\vspace*{-.2cm}
\includegraphics[width=0.38\textwidth]{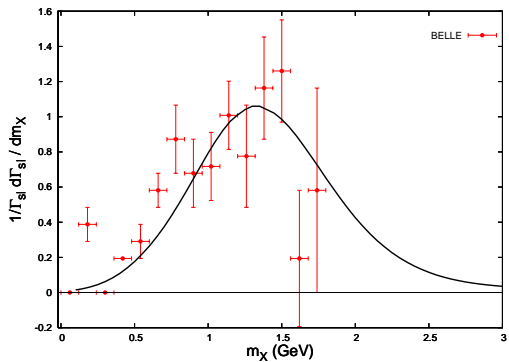}
\vspace*{-.1cm}
\caption{
\small
$B\to X_u\,l\,\nu_l$: $m_X$ 
distribution from Belle~\cite{belle}.
Data show the $\pi$ and the $\rho$ peak.
}
\label{slmxbel}
\vspace*{-.1cm}
\end{figure}
\begin{figure}[h]
\vspace*{-.2cm}
\includegraphics[width=0.38\textwidth]{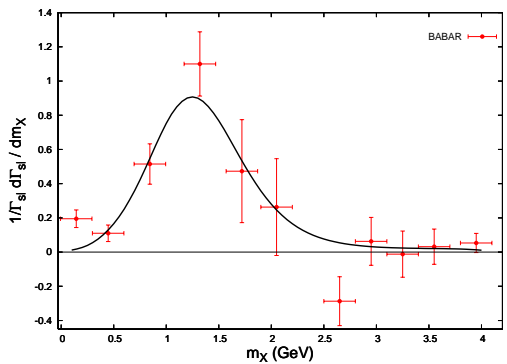}
\vspace*{-.1cm}
\caption{
\small
$B\to X_u\,l\,\nu_l$: $m_X$ 
distribution from BaBar~\cite{babar}.
Due to the larger binning only the $\pi$ peak is visible.
}
\label{slmxbab}
\end{figure}
\\
model the Doppler effect, due to the 
motion of the $B$ mesons  in the $\Upsilon(4S)$ 
rest frame.
The over-all agreement of the model with the data is
good for what concerns all the distributions in the radiative decay
and the $m_X$ distributions in semileptonic decays
in the region $m_X  >  1$ GeV, below which single resonances 
are expected to have a substantial effect in the dynamics.
The extracted values of $\alpha_S(m_Z)$ are in agreement with the world average
at most within two standard deviations (see Tab.~\ref{amz}).
\begin{figure}[h]
\vspace*{-.0cm}
\includegraphics[width=0.38\textwidth]{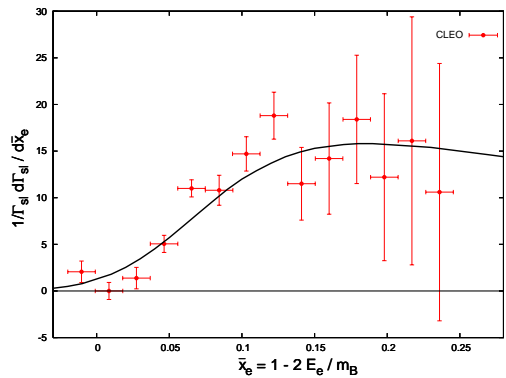}
\vspace*{-.1cm}
\caption{
\small
$B\to X_u\,l\,\nu_l$: electron energy 
distribution 
from CLEO~\cite{cleo}. 
}
\label{eecleo}
\vspace*{-.1cm}
\end{figure}
\begin{figure}[h]
\vspace*{-.0cm}
\includegraphics[width=0.38\textwidth]{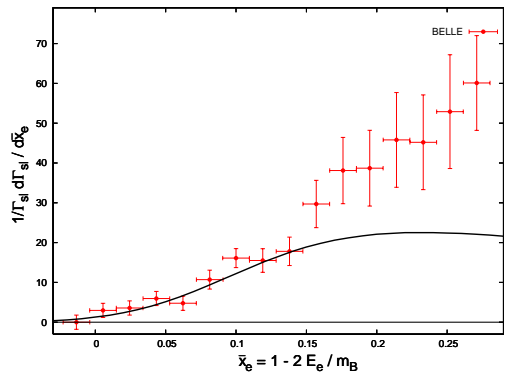}
\vspace*{-.1cm}
\caption{
\small
$B\to X_u\,l\,\nu_l$: electron energy 
distribution 
from Belle.
}
\label{eebelle}
\vspace*{-.0cm}
\end{figure}
\begin{figure}[h]
\vspace*{-.0cm}
\includegraphics[width=0.38\textwidth]{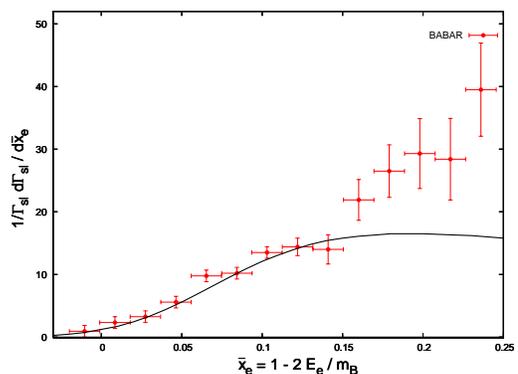}
\vspace*{-.1cm}
\caption{
\small
$B\to X_u\,l\,\nu_l$:  electron energy 
distribution 
from BaBar.
}
\label{eebabar}
\vspace*{-.0cm}
\end{figure}
\begin{figure}[ht]
\vspace*{-.0cm}
\includegraphics[width=0.4\textwidth]{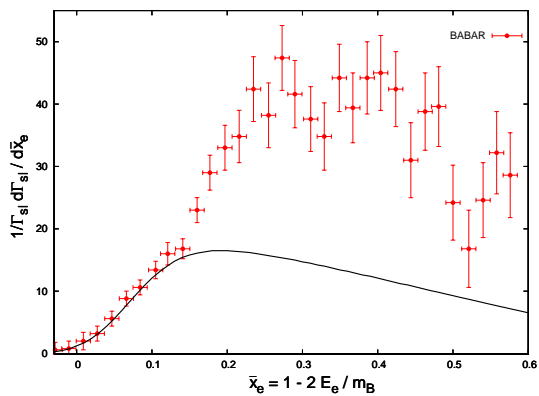}
\vspace*{-.0cm}
\caption{
\small
$B\to X_u\,l\,\nu_l$: electron energy 
distribution 
from BaBar (preliminary, error bars indicate statistical errors only).
}
\label{eecompl}
\end{figure}
\begin{table}
\vspace*{-.0cm}
\caption{\label{amz}
Extracted value of $\alpha_S(m_Z)$ compared with the PDG world average.
}
\begin{ruledtabular}
\begin{tabular}{l|l|l}
~~~~~~~~~~Spectrum &$\alpha_S(m_Z)$~~~~~~~~ & Error~~~~ \\
\hline
$ E_{\gamma} \quad  B \to X_s \gamma\quad$ CLEO& ~0.117 &  ~0.004\\
$E_{\gamma} \quad  B \to X_s \gamma\quad$ BaBar & ~0.129& ~0.005\\
$E_{\gamma} \quad  B \to X_s \gamma\quad$ Belle & ~0.130& ~0.005\\
$m_{\!X}\hspace{.24cm} B \to X_u l \nu_l\quad\!\!\!\!$ BaBar& ~0.119& ~0.003\\
$m_{\!X} \hspace{.24cm}B \to X_u l \nu_l\quad\!\!\!\!$ Belle& ~0.119& ~0.004\\
$E_e \quad B \to X_u l \nu_l\quad\!\!\!\!$ CLEO& ~0.117& ~0.005\\
$E_e \quad B \to X_u l \nu_l\quad\!\!\!\!$ BaBar& ~0.119& ~0.005\\
~~~~~~~~~~~PDG & ~0.1176& ~0.0020\\
\end{tabular}
\end{ruledtabular}
\vspace*{-.0cm}
\end{table}

The theory-data agreement is less clear in the case of the electron spectra
in semileptonic decay.
The agreement is acceptable in the charm background free region,
i.e. for $2.31~GeV \, < \, E_e \, < \, 2.64~GeV$.
There is not a good agreement with the preliminary BaBar spectrum for small
electron energies: 
our model predicts a broad maximum around $E_e \, = \, 2.1$ GeV,
while the data seem to peak at lower energies.
We do not known whether this discrepancy is related to a deficiency of our
model or to an under-estimated charm background.

\section{Semileptonic charmed $B$ decays}

To describe the semileptonic charmed $B$ decays $B\to X_c\,l\, \nu_l$
we need a formalism to take in account the (non negligible) charm mass $m_c$.
Standard threshold resummation has recently been  
generalized to the case of jets initiated by massive quarks ~\cite{resummass}. 
The inclusion of mass terms, in the 
$N$-moment space, results in a universal correction factor $\delta_N(Q^2;m^2)$
\vspace*{-.0cm}
\begin{eqnarray}
\label{maineq}
\!\!\!&\log&\!\!\! \delta_N(Q^2; m^2) =
\int_0^1 \!\!\frac{dy}y [(1-y)^{r(N-1)}-1]\\\nonumber
\!\!\!\!\!\!\!&\times&\!\!\!\!\!
\Bigg\{\!\!\!
 - \!\!\!\int_{ m^2 y^2 }^{ m^2 y } 
\!\!\frac{dk_{\perp}^2}{k_{\perp}^2} 
A[\alpha_S(k_{\perp}^2)]
\!-\!B[\alpha_S( m^2 y)]
\!+\!  D[\alpha_S( m^2 y^2)]
\!\!\Bigg\},
\vspace*{-.0cm}
\end{eqnarray}

\noindent
so that the QCD form factor for massive quarks reads
\begin{equation}
\label{univ}
\sigma_N(Q^2;  m^2)  =  \sigma_N(Q^2) ~ \delta_N(Q^2;  m^2),
\vspace*{-.0cm}
\end{equation}
where $\sigma_N(Q^2)\equiv \sigma_N(Q^2;m^2\!\!=\!\!0)$ 
is the standard QCD form factor for a massless quark
defined in Eq.~\ref{expoGN} and $r\equiv m^2/Q^2\ll 1$ 
is the corrections mass parameters. 

Combining the above resummed formula with the full $O(\alpha_S)$ triple
differential distribution recently computed \cite{agru}, \cite{trott}
and using the model coupling described in the previous section,
we can analyze experimental data from semileptonic $b\to c$ transitions.

\section{Conclusions}
\label{section6}

We have presented a model for the QCD form factor describing
radiative and semileptonic $B$ decay spectra, based on
soft-gluon resummation to NNLL accuracy and on a power expansion 
in an analytic time-like coupling.
The latter is free from Landau singularities 
and resumes absorptive effects in gluon cascades to all orders.

The agreement with invariant hadron mass distributions in radiative
and semileptonic decays measured by CLEO, BaBar and Belle is pretty good. 
The extracted values of $\alpha_S(m_Z)$ are in agreement with the current
PDG average within  at most two standard deviations (see Tab.~\ref{amz}).

The agreement with the electron spectra in semileptonic decays is instead, 
in general, not so good. 
At present, we do not know whether the discrepancy is 
due to a deficiency of our model or to an under-subtracted
charm background. 
Let us stress however than non-perturbative effects are expected to
by much smaller in the electron spectrum than in the case of the
other analyzed spectra \cite{noi}.

Using a new resummation formalism recently developed together 
with
available fixed order calculation, we will
carry out a phenomenological analysis of semileptonic charmed decays soon.


\begin{thebibliography}{99}

\bibitem{cabibbo}
  N.\ Cabibbo, Phys.\ Rev.\ Lett.\ {\bf 10}, 531 (1963).

\bibitem{km}
  M.\ Kobaya\-shi and T.~Maskawa, Prog.\ Theor.\ Phys.\ {\bf 49}, 652 (1973).

\bibitem{uraltsev}
 I.~I.~Y.~Bigi, M.~A.~Shifman, N.~G.~Uraltsev and A.~I.~Vainshtein,
  Int.\ J.\ Mod.\ Phys.\  A {\bf 9} (1994) 2467
  [arXiv:hep-ph/9312359].

\bibitem{ugo2001}
  U.\ Aglietti,
  Nucl.\ Phys.\  B {\bf 610} (2001) 293
  [arXiv:hep-ph/0104020];
  Nucl.\ Phys.\ Proc.\ Suppl.\  {\bf 157} (2006) 141
  [arXiv:hep-ph/0601242].

\bibitem{noi}
  U.\ Aglietti, G.\ Ricciardi and G.\ Ferrera,
   Phys.\ Rev.\  D {\bf 74} (2006) 034004
  [arXiv:hep-ph/0507285];
  Phys.\ Rev.\  D {\bf 74} (2006) 034005
  [arXiv:hep-ph/0509095];
  Phys.\ Rev.\  D {\bf 74} (2006) 034006
  [arXiv:hep-ph/0509271].

\bibitem{cattren}
  S.~Catani and L.~Trentadue,
  Nucl.\ Phys.\ B {\bf 327} (1989) 323.
\bibitem{sterman}
  G.~Sterman,
  Nucl.\ Phys.\ B {\bf 281} (1987) 310.

\bibitem{shirkov}
  D.\ V.\ Shirkov and I.\ L.\ Solovtsov,
  Phys.\ Rev.\ Lett.\  {\bf 79} (1997) 1209
  [arXiv:hep-ph/9704333].

\bibitem{stefanis}
  A.~I.~Karanikas and N.~G.~Stefanis,
  Phys.\ Lett.\  B {\bf 504} (2001) 225
  [Erratum-ibid.\  B {\bf 636} (2006) 330]
  [arXiv:hep-ph/0101031].

\bibitem{ugogiu2004}
  U.~Aglietti and G.~Ricciardi,
  Phys.\ Rev.\  D {\bf 70} (2004) 114008
  [arXiv:hep-ph/0407225];
 U.~Aglietti, G.~Ferrera and G.~Ricciardi,
  Nucl.\ Phys.\  B {\bf 768} (2007) 85
  [arXiv:hep-ph/0608047].
 .

\bibitem{mp}
  S.~Catani, M.~L.~Mangano, P.~Nason and L.~Trentadue,
Phys.\ Lett.\  B {\bf 378} (1996) 329
  [arXiv:hep-ph/9602208].

\bibitem{gardi}
E.~Gardi,
  JHEP {\bf 0404} (2004) 049
  [arXiv:hep-ph/0403249];
  J.~R.~Andersen and E.~Gardi,
  JHEP {\bf 0601} (2006) 097
  [arXiv:hep-ph/0509360].

\bibitem{neubert}
  S.~W.~Bosch, B.~O.~Lange, M.~Neubert and G.~Paz,
  Nucl.\ Phys.\  B {\bf 699} (2004) 335
  [arXiv:hep-ph/0402094];
 B.~O.~Lange, M.~Neubert and G.~Paz,
  Phys.\ Rev.\  D {\bf 72} (2005) 073006
  [arXiv:hep-ph/0504071].
\bibitem{ligeti}
  C.~W.~Bauer, Z.~Ligeti and M.~E.~Luke,
  Phys.\ Rev.\  D {\bf 64} (2001) 113004
  [arXiv:hep-ph/0107074].

\bibitem{modfrag}
  U.~Aglietti, G.~Corcella and G.~Ferrera,
  Nucl.\ Phys.\ B (2007), doi: 10.1016/j.nuclphysb.2007.04.014,
 [arXiv:hep-ph/0610035].

\bibitem{babar}
  B.~Aubert {\it et al.}  [BABAR Collaboration],
  arXiv:hep-ex/0408068; arXiv:hep-ex/0507001; 
 Phys.\ Rev.\  D {\bf 72} (2005) 052004
  [arXiv:hep-ex/0508004];
 Phys.\ Rev.\  D {\bf 73} (2006) 012006
  [arXiv:hep-ex/0509040].

\bibitem{belle}
  P.~Koppenburg {\it et al.}  [Belle Collaboration],
  Phys.\ Rev.\ Lett.\  {\bf 93} (2004) 061803
  [arXiv:hep-ex/0403004];
 A.~Limosani {\it et al.}  [Belle Collaboration],
  Phys.\ Lett.\  B {\bf 621} (2005) 28
  [arXiv:hep-ex/0504046];
 I.~Bizjak {\it et al.}  [Belle Collaboration],
  Phys.\ Rev.\ Lett.\  {\bf 95} (2005) 241801
  [arXiv:hep-ex/0505088].


\bibitem{cleo}
  S.~Chen {\it et al.}  [CLEO Collaboration],
  Phys.\ Rev.\ Lett.\  {\bf 87} (2001) 251807
  [arXiv:hep-ex/0108032].

\bibitem{resummass}
  U.~Aglietti, L.~Di Giustino, G.~Ferrera and L.~Trentadue, 
[arXiv:hep-ph/0612073].
\bibitem{agru}
V.~Aquila, P.~Gambino, G.~Ridolfi and N.\,\,Uraltsev,
Nucl.\ Phys.\ B {\bf 719} (2005) 77
[arXiv:hep-ph/0503083].
\bibitem{trott}
M.~Trott,
Phys.\ Rev.\ D {\bf 70} (2004) 073003
[arXiv:hep-ph/0402120].

\end{thebibliography}
\end{document}